# 'Conspiracy Machines' - The Role of Social Bots during the COVID-19 'Infodemic'

## Research-in-progress

### Julian Marx
Department of Computer Science and Applied Cognitive Science
University of Duisburg-Essen
Duisburg, Germany
Email: julian.marx@uni-due.de

### Felix Brünker
Department of Computer Science and Applied Cognitive Science
University of Duisburg-Essen
Duisburg, Germany
Email: felix.bruenker@uni-due.de

### Milad Mirbabaie
Faculty of Business Studies and Economics
University of Bremen
Bremen, Germany
Email: milad.mirbabaie@uni-bremen.de

### Eric Hochstrate
Department of Computer Science and Applied Cognitive Science
University of Duisburg-Essen
Duisburg, Germany
Email: eric.hochstrate@uni-due.de

## Abstract

The omnipresent COVID-19 pandemic gave rise to a parallel spreading of misinformation, also referred to as an 'Infodemic'. Consequently, social media have become targets for the application of social bots, that is, algorithms that mimic human behaviour. Their ability to exert influence on social media can be exploited by amplifying misinformation, rumours, or conspiracy theories which might be harmful to society and the mastery of the pandemic. By applying social bot detection and content analysis techniques, this study aims to determine the extent to which social bots interfere with COVID-19 discussions on Twitter. A total of 78 presumptive bots were detected within a sample of 542,345 users. The analysis revealed that bot-like users who disseminate misinformation, at the same time, intersperse news from renowned sources. The findings of this research provide implications for improved bot detection and managing potential threats through social bots during ongoing and future crises.

# 1   Introduction

Social media have become a powerful communication tool for the procurement and dissemination of information (Varol et al. 2017). Especially during extreme events, social media are used to engage with online contacts and to stay informed (Mirbabaie et al. 2020; Mirbabaie and Marx 2019). The same applies to the infectious Coronavirus disease 2019 (COVID-19), which turned out to be one of the most discussed events on social media in recent years. However, apart from human users, social bots, that is, algorithms programmed to mimic human behaviour on social media platforms, also have been found engage in online discussions (Stieglitz, Brachten, et al. 2017). Scholarship distinguishes between benign social bots, which were designed with good intentions, and malicious bots. The aim behind malicious social bots may include altering other humans' behaviour or distorting the opinion climate in social media communities (Ferrara et al. 2014; Ross et al. 2019). However, examples in the recent past have shown that, without proper monitoring, also benign bots can turn malicious. For instance, in 2016, Microsoft's Tay had to be shut down due to obscene and inflammatory tweets it had learned from other Twitter users (Neff and Nagy 2016). Furthermore, the unregulated nature of social media entails the dissemination of information of questionable credibility (Wattal et al. 2010). The spread of misinformation, defined as "*false or inaccurate information, especially that which is deliberately intended to deceive*" (Lazer et al. 2018), poses a serious threat to public health during pandemics (Zarocostas 2020). The amplification of the spreading of misinformation by social bots becomes problematic if the health information and contributions, for example, contain information that can influence the public opinion in a wrong way and thereby impede public health measures.

The distribution of health information, which is not consistent with information from evidence-based sources, is a problem that emerged at the beginning of this century (Eysenbach 2002; Kim and Dennis 2019). Since then, the quantity and dissemination of misinformation have grown in unprecedented ways (Shahi et al. 2020). Information epidemiology, or infodemiology, is an emerging research field, which focuses on the determinants and distribution of health information in an electronic medium and specifically the internet. It aims to identify areas of knowledge translation gaps between the best evidence provided by experts and what most people believe (Eysenbach 2002). Research in this area remains crucial since the diffused health information can alter the effectiveness of the countermeasures taken by the government and also spread hysteria which can have a negative impact on the mental well-being of social media users (Rosenberg et al. 2020). Within the information systems discipline, a debate has been launched that questions the ability of IS research to actually help fight pandemics such as COVID-19. Scholarship that contributes to infodemiology and provides knowledge about 'infodemics' containment, we argue, can make an important contribution. Thus, in this study, we focus on the detection of social bots with a malicious intent, i.e. spreading rumors and unverified information. The research is guided by the following research question:

**RQ:** *To what extend do social bots affect the Twitter discussion about COVID-19 by disseminating misinformation?*

In order to answer this research question, we apply social bot detection techniques based on three metrics: *Tweet Uniqueness (TU)*, *Tweet Frequency (TFQ)*, and *Friends-Followers-Ratio (FFR)*. The underlying data set covers a period of 12 weeks of twitter communication about the COVID-19 pandemic and includes 3,052,030 tweets authored by 542,345 users. Subsequently to the bot detection, a manual content analysis was conducted to identify misinformation among the social bot content. Consequently, this study aims to contribute to scholarship on infodemiology by characterizing social bots and their role during an 'infodemic'. We further provide a basis of discussion about social bot detection and provide implications for practitioners to better control malicious social bot content.

In the remainder, we provide a literature background, explain our methical approach and present preliminary results. To conclude this paper, the next steps of this research-in-progress are presented.

# 2   Related Work

## 2.1   Infodemiology

The term infodemiology is a portmanteau of information and epidemiology and can be defined as the science of distribution and determinants of disease in populations and information in an electronic medium, specifically the Internet and social media (Eysenbach, 2009). The latter is defined as "*a group of Internet-based applications that build on the ideological and technological foundations of Web 2.0, and that allow the creation and exchange of User Generated Content.*" (Kaplan and Haenlein 2010). It has the ultimate aim to provide public health professionals, researchers, and policy makers with the required tools and information to influence public health and policy decisions (Eysenbach 2002, 2009). This research area initially focused on the identification of misinformation. Since then, the term has been used to analyse the relationship between the demand for health information, which can be investigated for example through web queries analysis, and the supply of health information, for which supply-based applications are carried out like social media analysis (Zeraatkar and Ahmadi 2018).

Metrics and approaches of infodemiology are an integral part of health informatics, the most popular sources being Twitter and Google. The potential and feasibility of using social media to conduct studies with an infodemiological approach has been demonstrated in previous studies (Mavragani 2020; Mavragani et al. 2018). Eysenbach (2002), who first proposed the term 'infodemiology', suggested during the SARS pandemic that the use of population health technologies, such as the Internet, can help detect outbreaks of the disease at an early stage. During the H1N1 pandemic, an infodemiological approach was used called 'infoveillance', which describes the usage of infodemiology methods for surveillance purposes. This study illustrated the potential of using social media to conduct infodemiological studies and showed that H1N1-related tweets on Twitter were primarily used for the dissemination of information from credible sources, but also included many opinions and experiences (Chew and Eysenbach 2010). Another infoveillance study on this subject matter identified main topics that were most discussed by Twitter users and additionally analysed the sentiments of the tweets (Abd-Alrazaq et al. 2020).

## 2.2 Social Bots

Typically, crises communication on social media is dominated by customary actors such as individuals from the general public, affected organisations, and influential individuals such as politicians, journalists, and influencers (Mirbabaie and Zapatka 2017; Stieglitz, Bunker, et al. 2017). Emerging actors in this context are social bots (Brachten et al. 2018). The identification of bot accounts is of increasing interest in research and society, especially on Twitter (Bruns et al. 2018). Recent studies have been conducted to describe how bots are used to amplify messages on social media (Brünker et al. 2020; Vosoughi et al. 2018) or how fake content is populated on Twitter (Gupta et al. 2013). In general, social bots are automated accounts that exhibit human-like behaviour with either malicious or benign objectives (Ferrara et al. 2014). The detection of social bots has become more complex because social bots become increasingly sophisticated. In this respect, it has become more difficult for individuals to distinguish between a human and a social bot account (Freitas et al. 2015). Considering that between 9% and 15% of active Twitter accounts might be bots (Varol et al. 2017), the detection of social bots on social media poses a fundamental challenge.

Recent work has also focused on how social bots can bias discussions within social networks (Ferrara et al. 2014; Ross et al. 2019). For example, social bots successfully influenced political discussions and even had an effect on the outcomes of elections (Brachten et al. 2017; Ferrara 2017). Social bots are also applied for spamming and thereby diverting attention from discussed issues (Bradshaw and Howard 2017) or overstating trends (Brachten et al. 2018). For this purpose, social bot strategies like smoke screening, astroturfing, and misdirecting are used (Stieglitz, Brachten, et al. 2017). Messias et al. (2013) and Zhang et al. (2013) show that social bots can cooperate to manipulate the influence scores of several centrality measures. Twitter, for instance, already took countermeasures against bots with the goal of reducing the impact of such accounts' malicious actions and therefore targeted accounts that had been actively used to amplify and disseminate news of questionable sources (Onuchowska et al. 2019). However, with unprecedented amounts of communication and emerging phenomena such as the COVID-19 'infodemic', it remains imperative to conduct research that aims at understanding the technological enablers and misinformation mechanisms behind social bot activity.

# 3 Research Design

## 3.1 Case Description

After the first appearance of the infectious respiratory coronavirus disease in Wuhan, China, it vastly spread in China and other countries around the world (World Health Organization 2020a). The disease is also known as COVID-19 and causer of a global crisis. For Germany, the virus has not been as unpredictable as other emerging infectious diseases because in Germany the first cases have been confirmed on January 28, 2020 while in China the first cases have been confirmed on January 11 (World Health Organization 2020b). The opportunity for individuals to share their own stories, feelings, opinions, judgments, or evaluations about the virus on social media channels did not just lead to an increase of individual activity but also social bot activity on online social platforms such as Twitter, Facebook, and YouTube (Abd-Alrazaq et al. 2020; Laato et al. 2020). To further investigate the spread of misinformation, an infodemiological approach is going to be conducted, where the idea of the field is to measure the pulse of public opinion, attention, behaviour, knowledge and attitudes by tracking what people do and write on the Internet (Eysenbach 2009).

## 3.2 Data Collection

The data used in this study was collected via the Twitter API which is a commonly used tracking method since it enables researchers to retrieve a wide range of meta-data (Stieglitz, Mirbabaie, *et al.*, 2018). Overall, data was crawled during the time period of February 27 (0:00 UTC) to May 20, 2020 (23:59 UTC) by applying a self-developed Java crawler using the Twitter4J1 library and saving it in a MySQL database. In general, three separate crawling processes were conducted because new relevant hashtags and terms appeared in the public discourse over the course of time, for example, after the coronavirus got an official name by the WHO in February. The first crawling process tracked the following terms: Coronavirus, nCoV2019, WuhanCoronaVirus, WuhanVirus, CoronaVirusOutbreak, Ncov2020 and coronaviruschina. To retrieve a reliable dataset, a second and third data crawling were conducted by tracking tweets using at least one of the terms Covid19, covid-19, covid_19, sarscov2, covid, cov, corona, corona Germany, #COVID19de, #coronapocalypse, #staythefuckhome, #flattenthecurve, and #stopthespread. The third data crawling contained specific terms which were especially popular in German-speaking tweets. Finally, all three crawled data sets were merged, and duplicates were removed, representing the foundation for this research. The tracking yielded a total of 542,345 users creating 3,052,030 tweets. A directed retweet network for further analyses is created, consisting of users (vertices) and retweets (edges).

## 3.3 Social Bot Detection

Only highly active users are considered for the creation of a sample in this research because the activity is also reflected in the likelihood of the account to be a bot (Ferrara et al. 2014). To this end, accounts which posted at least 150 tweets within two weeks during the tracking period were considered highly active. Tweets, retweets and mentions are included in the dataset to cover the entire spectrum of Twitter communication. For the identification of bot-like accounts, the dataset was preprocessed, i.e. deletion of unnecessary meta data and exclusion of inactive users. The social bot identification as well as the preprocessing were performed by using R which is a free software environment for statistical computing and graphics (Nordhausen 2015). Three metrics are considered for the identification of the most social bot-like users: (1) the Tweet Uniqueness (TU), (2) the Tweet Frequency (TFQ), and (3) the Friends-Follower-Ratio (FFR) and for every metric, a specific threshold was derived to create a sample of potential social bots (Brünker et al., 2020).

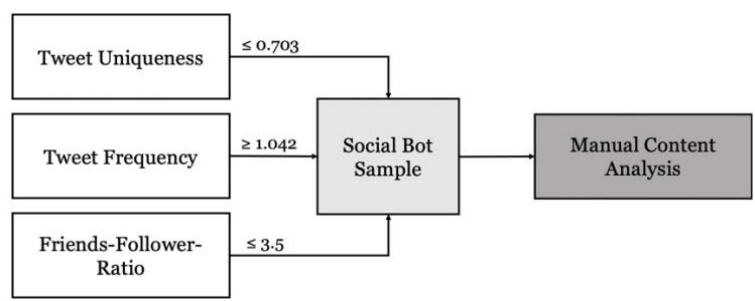

*Figure 1.    Social Bot Sampling Procedure (adapted by Brünker et al., 2020).*

The TU quantifies the particularity of the content of a tweet. For the computation of this measure, the number of distinct tweets is divided by the total number of tweets *k* by the user, according to (Ross et al. 2018). A tweet is unique if the tweet alters in at least one character from every other user's tweet. The TU will equal 1 if every tweet by a user is different from others and the more identical tweets a user publishes the more this value decreases. The global TU in the dataset, which is the number of unique tweets divided by the total number of extracted tweets, serves as the threshold every user must fall below in order to be classified as a potential bot (82,692/117,678 = 0.703).

A high tweet frequency is and indicator for bot-like behaviour (Brachten et al. 2017; Varol et al. 2017). The TFQ displays the activity of a user during the tracking period and is calculated by dividing the total number of tweets of the user by the number of hours of the tracking period. For example, a TFQ of 3 means that the user publishes three tweets per hour each day. In this regard, the threshold of 1.042 must be exceeded in order for the user to be assigned to the social bot sample. In this case it means that a user must have disseminated at least 25 tweets a day during the tracking period.

The FFR indicates the relation between the friends and the followers of a user. In recent studies, the relation between a users' friends and followers was considered for social bot detection (Chu et al. 2012; Ferrara et al. 2014). The relationships on Twitter are unidirectional, i.e. if a user *A* follows the user *B*, then *A* is a follower of *B* and *A's* friend is *B*. The ratio is calculated by dividing the total number of friends by the total number followers of a user and the result must exceed the threshold of a ratio of 3.5. Calculating a global FFR resulted in a way too low value which has led to a congested social bot sample because most of the active accounts have been assigned to the sample. The minimum FFR was set 3.5 so that, for example accounts with 1,000 followers must be friends with at least 3,500 other users in order to be classified as potential bots.

## 4 Preliminary Results

Overall, 380 accounts, who are responsible for 117,678 tweets, are considered highly active within the dataset. 20,5% (78 accounts) of them meet at least one of the conditions and are therefore classified as likely social bots. These, in turn, disseminated 19,117 tweets throughout a 12-week time period. The manual content analysis has been conducted for the 78 classified accounts.

Table 1 shows selected examples of tweets by some of the alleged bots. The tweets contain misinformation or conspiracy content, but the mentioned users did not disseminate this kind of content solely. In particular, there are also retweets of news and updates about the virus. None of the accounts in the social bot sample could be classified as a social bot based on a single tweet. However, analysing the account's characteristics according the metrics TU, TFQ or FFR allows to better classify the distinct accounts. Particularly some accounts significantly surpassing the threshold of the metric TFQ. For instance, one account created 3,967 tweets, including news, many hashtags, and links to related websites or retweeted his own tweets.

| User | Tweet | #Retweets |
|---|---|---|
| @C***64K**z | RT @WaltiSiegrist: The worldwide biological weapons research of the USA #SARSCoV2 #coronavirus #covid19 | 20 |
| @F***R*5 | RT @lobbycontrol: Did #party donations play a role in the occupation of the #NRW-#Corona Council? More than 1.1 million Euro flowed 2002-2017 from the company #Trumpf and its owners to #CDU (765.766 €) and #FDP (335.000 €). Trumpf is represented on the council by the billionaire Nicola #Leibinger-Kammüller | 67 |
| @L****iB | RT @hfeldwisch: How #Taiwan prevented the #COVID19 outbreak in his country - and the @WHO doesn't want to know about it: Exciting article by Richard Friebe @TspWissenschaft | 0 |

*Table 1.   Selected examples of social bots spreading misinformation or conspiracy theories.*

## 5 Conclusion and Next Steps

The aim of this research is to determine the extent of the impact social bots have on the network by disseminating misinformation or conspiracy theories and to provide an evaluation on whether the identified social bots can be considered a threat measured by their influence. Particularly, in this global pandemic, social bots have been neglected so far and the objective of this research is to gain new insights into how social bots are applied in this context. For this purpose, this research-in-progess carried out the first steps of data collection, preprocessing and manual analyzing.

The manual assessment of bot-like accounts that entered the sample as they passed over the thresholds defined by Brünker et al. (2020) revealed that their impact on the COVID-19 'infodemic' is very limited. However, the analysis provides valuable insights that help us to better understand online behaviour that is relevant to an 'infodemics' context. Those mechanisms do not only apply to social bots but can be transferred onto other social media users trying to exert influence within an 'infodemic' environment.

However, the number of relevant tweets grew during the data acquisition process. As a result, some of the tweets among the hashtags examined were not tracked in time and no absolute completeness of the data set can be guaranteed. To properly analyse the high number of tweets in our dataset, an automated content analysis may be applied as part of the future work. Since "automated accounts are particularly active in the early spreading phases of viral claims" (Shao *et al.*, 2017, p. 11) and there is missing data between the first infection and the beginning of the data crawling, namely January 11 to February 27, it is possible that many social bots were not caught or already blocked or deleted by Twitter itself.

To analyse the impact of these social bot accounts on the dissemination of misinformation and conspiracy theories, their individual PageRank value within a retweet network can be calculated and compared. As a next step, the PageRank score can be evaluated within a created retweet network of the identified accounts. PageRank is a variation of the Eigenvector centrality measure which considers network information, i.e. users who are closely related to influential users, to quantify the influence of a node (Bonacich 1972). It is developed for directed graphs such as retweet networks and integrates the flow of influence to calculate the authority of a node in a network (Yang and Tang 2012). The PageRank scores of the social bots can be compared to other users' scores and on the basis of this, statements about the impact of the social bots throughout the network can be derived.

Moreover, the tweets of social bots are likely to be retweeted because misinformation spreads faster than information from credible sources (Shu et al. 2020). Also, social bots can cooperate and thereby reach a higher PageRank value than other users (Messias et al. 2013; Zhang et al. 2013). In this case and additionally, a tweet of a social bot is retweeted by other influential users, the bot is very likely to be considered influential as well. Depending on the results of the PageRank algorithm and the identified bots being influential or not, they can be classified as threat or harmless.

## Acknowledgements

This project has received funding from the European Union's Horizon 2020 research and innovation programme under the Marie Skłodowska-Curie grant agreement No 823866.